\documentstyle[prl,aps,graphicx,amssymb]{revtex}
\renewcommand{\H}{{\cal H}}
\renewcommand{\c}{{\bf c}}
\draft
\begin{document}
\twocolumn[\hsize\textwidth\columnwidth\hsize\csname
@twocolumnfalse\endcsname

\title{Correlated random networks}
\author{Johannes Berg and Michael L\"assig}
\address{ Institut f\"ur Theoretische Physik,
Universit\"at zu K\"oln\\
Z\"ulpicher Stra{\ss}e 77,
50937 K\"oln,
Germany}

\date{\today}
\maketitle

\begin{abstract}
We develop a statistical theory of networks.
A network is a set of vertices and links given by its
adjacency matrix $\c$, and the relevant statistical ensembles
are defined in terms of a partition
function $Z=\sum_{\c} \exp {[}-\beta \H(\c) {]}$. The
simplest cases
are uncorrelated random networks such as the well-known
Erd\"os-R\'eny graphs. Here we study more general interactions
$\H(\c)$ which lead to {\em correlations}, for example, between
the connectivities of adjacent vertices. In particular, such
correlations occur in {\em optimized} networks described by
partition functions in the limit $\beta \to \infty$. They
are argued to be a crucial signature of evolutionary design
in biological networks.
\end{abstract}

\vskip2pc
]
\narrowtext

Networks describe structures as diverse as the interaction
links between proteins in a cell, the wiring of the brain,
or the connections of the internet. Recent theoretical work
\cite{barabasi_review,dorogovtsev_review}
has focused on communication networks, and a wealth of
quantitative data is now becoming available on networks in
molecular biology. Examples include control networks in
gene transcription \cite{gene_network}, the interaction map of
proteins \cite{protein_network}, and the pathways of cell metabolism
\cite{metabolic_network}. All these systems consist of many
different kinds of molecules linked by complex interactions.
Network models are a simplified description, which neglects
quantitative aspects of these interactions and focuses
solely on their pathways.

A network is a set of vertices $i = 1,
\dots, N$ connected by links. It is uniquely defined
by the adjacency matrix $\c$, whose entries are $c_{ij} = 1$
if there is a link from $i$ to $j$ and $c_{ij} = 0$
otherwise. We consider here networks with undirected links,
where $\c$ is symmetric. The {\em connectivity} or {\em degree} of a
vertex
is then defined as the total number of links connected to
it, $k_i \equiv \sum_j c_{ij}$. The {\em distance} $d_{ij}$
between two vertices $i$ and $j$ is the number of links along the
shortest path connecting them \cite{bollobas}. We assume the vertices are
labeled (for example, by their biochemical identity),
so that the correspondence between adjacency
matrix $\c$ and its graph is one-to-one~\cite{unlabel}.

Networks with an irregular wiring naturally lend themselves
to a statistical description \cite{bollobas}. We discuss here the
equilibrium statistics of networks. 
The partition function $Z$ can be 
defined as a sum over all graphs with a fixed number 
$N$ of vertices and a  fixed number 
$M=\sum_{i<j} c_{ij}={\rm Tr} \, \c^2/2$  
of links
\begin{equation}
\label{partition1}
Z=\prod_{i<j} \sum_{c_{ij}=0}^1 
         \delta(M- {\rm Tr} \, \c^2/2) \, 
         \exp[-\beta \H(\c) ] \;.
\end{equation}
Averages over this ensemble are denoted by $\langle \dots
\rangle$. 
Alternatively one can define $Z$ with an arbitrary 
number of links adjusted by a suitable chemical potential. 
The ensembles of relevance here have a finite average connectivity
$\kappa \equiv 2 M / N$. For fixed $\kappa$, the distribution of
connectivities, 
\begin{equation}
p(k) \equiv  \frac{1}{N}
             \sum_{i=1}^N \langle \delta(k_i - k) \rangle \;,
\label{pk}
\end{equation}
becomes asymptotically independent of $N$, implying that
typical adjacency matrices $\c$ become sparse for large~$N$.
This is the case of interest for applications. 

A satisfactory mathematical theory exists to date only
for what we call {\em uncorrelated random networks}
\cite{newman,burda}. 
In this
case, the Hamiltonian depends only on single-point connectivities,
\begin{equation}
\H_1 (\c) = \sum_{i=1}^N f (k_i) \;,
\label{H1}
\end{equation}
leading to
$p(k) \sim \exp[-\beta f(k) - \mu k]/k! $ where the constant 
of proportionality is fixed by normalization and $\mu$ is adjusted to 
give the correct average connectivity. 
Since all graphs with the same connectivities $k_1, \dots,
k_N$ have the same statistical weight, this ensemble ensures
the maximally random wiring compatible with the
distribution $p(k)$. The simplest example is the well-known
Erd\"os-R\'eny graphs, where $\beta \H = 0$ and $p(k)$ is a
Poissonian.

Many natural networks are, however, not of this type.
The simplest kind of correlations
occur if the joint
distribution of connectivities for neighboring vertices,
\begin{equation}
q(k,k') \equiv \frac{1}{\kappa N}
     \sum_{i,j=1}^N \langle \delta(k_i - k) c_{ij}
                            \delta(k_j - k') \rangle \;,
\label{q}
\end{equation}
differs from its form for uncorrelated random networks,
$q_0 (k,k') = (k k'/\kappa^2) p(k) p(k')$. Higher
correlations can be defined in a similar way~\cite{higher}.
Connectivity correlations have been found in growth
models of communication networks \cite{newman_grow,vespignani}
as well as in data
of genetic and protein networks\cite{alon,maslov}.

These observations call for a statistical theory of more general
ensembles called {\em correlated random networks}, which is
the subject of this Letter. The ensembles of interest are
characterized by finite distributions $p(k)$ and $q(k,k')$
in the limit of large $N$. One then expects a universal
logarithmic scaling of
the average distance $d_{ij}$ between vertices,
$\sum_{i,j} \langle d_{ij} \rangle / N_\Omega^2 \sim \log N_\Omega$,
in any connected component $\Omega$ with $N_\Omega$ 
nodes~\cite{logscal}. This is consistent
with our numerical findings. Hence, correlated random
networks maintain a sparse connectivity matrix and are
locally tree-like.  The `inverse temperature' $\beta$ in
(\ref{partition1}) measures the
deviation from Erd\"os-R\'eny graphs. Quite remarkably,
these structural properties are preserved in the limit
$\beta \to \infty$, where we obtain nontrivial
{\em optimized} networks. Ensembles of this kind 
generically have strong correlations.

The simplest type of Hamiltonian producing correlations has
nearest-neighbor connectivity interactions,
\begin{equation}
\H_2 (\c)  = \sum_{i<j} c_{ij} g(k_i,k_j) \;,
\label{H2}
\end{equation}
where $g(k,k')$ is some function of the connectivities. 
The resulting class of graph ensembles can be seen as a showcase for
correlated random networks where analytic expressions
can be derived. Higher order correlations are generated by Hamiltonians 
with next-nearest neighbor interactions etc. We also 
study 
a Hamiltonian $\H = \H_1 + \lambda \H_d$ with a nonlocal part,
\begin{equation}
\H_d (\c) =  \sum_{i<j}^N d_{ij} \;,
\label{Hd}
\end{equation}
often called the {\em diameter} of the graph. For
$\lambda > 0$ and a suitable scaling $\lambda \sim 1/ (N \log
N)$, this Hamiltonian is found to generate
{\it compactified networks} with finite $p(k)$ and $q(k,k')$,
provided the 
extra term $\H_1$ stabilizes the network against collapse to
a star. 
\begin{figure}[t]
\includegraphics[width=.45 \linewidth,angle=90]
{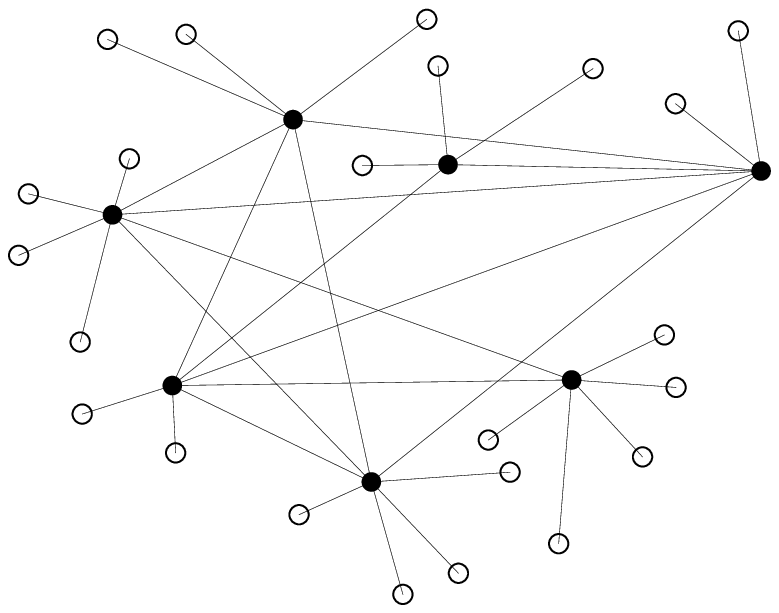}
\includegraphics[width=.45 \linewidth]
{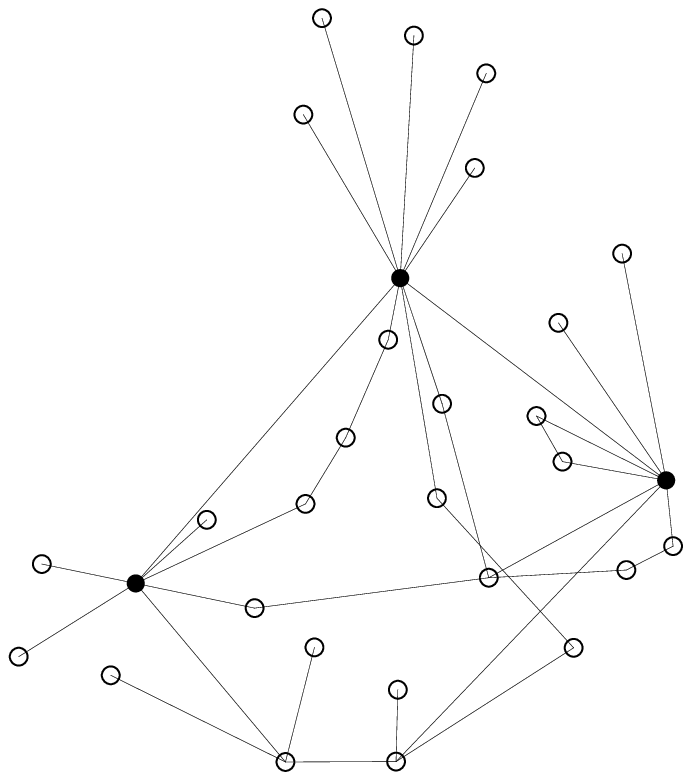}
\caption{\label{fig1} Optimized networks, obtained from local
interactions (left) and nonlocal
interactions for large
$\beta$. Hubs of high connectivity (filled circles) are
preferentially connected to peripheral vertices of low
connectivity (empty circles).}
\end{figure}
Compactified networks occur in communication
and transport~\cite{smallworld}, and may play a role in biology
\cite{jeong,sole}. For example, distance-optimized
networks obtained from (\ref{Hd}) show an abundance of
high-connectivity vertices (hubs) which are preferentially
connected to peripheral vertices of low connectivity. A
very similar structure (with even more pronounced hubs)
can also be obtained from local interactions of the form
(\ref{H2}). Typical graphs are shown in Fig~\ref{fig1}.

To establish these results, we first discuss the analytical
treatment of local interactions, choosing a generic
Hamiltonian of the form ${\cal H} = {\cal H}_1 + {\cal H}_2$
given by (\ref{H1}) and (\ref{H2}). From the partition
function (\ref{partition1}), the free energy  
is derived by using an integral
representation for the constraints due to the connectivity
$k_i$ at each vertex and making a Hubbard-Stratonovich
transformation. The resulting integral can be evaluated in
a saddle-point approximation, yielding the reduced free energy
per vertex in the thermodynamic limit,
\begin{eqnarray}
\label{ZQ}
 -&& \beta f  \equiv   \lim_{N\to\infty} \frac{1}{N}\log Z
 = \frac{\kappa}{2} \log N 
 +\frac{\kappa}{2} \left(\log \kappa -1\right)  \nonumber \\
&&-\frac{\kappa}{2} \log \left(\sum_{k,k'}Q_k 
{\cal G}^{-1}_{kk'}Q_{k'} +{\rm e}^{-2\mu} \right) \\
&&+\log \left[ \sum_k \frac{(Q_k+{\rm e}^{-\mu})^k}{k!} 
\exp\left[-\beta f(k)\right] \right]
\nonumber 
\end{eqnarray}
The `order parameters' $Q_k$ and the chemical potential $\mu$ 
have to be determined self-consistently from the saddle-point condition.
This form is closely related to a field-theoretic
approach~\cite{burda,krzywicki,tadpole}, where networks
appear as the Feynman diagrams of a Gaussian integral with 
a propagator matrix ${\cal G}_{kk'}=\exp[-\beta g(k , k')]-1$ and 
interactions as specified by the last term in (\ref{ZQ}). 
Notice the super-extensive scaling of the entropy,
$(\kappa/2) \log N$, which reflects that unlike in a regular
lattice, each vertex can be connected to all $N - 1$ other
vertices. The last term in (\ref{ZQ}) is directly related
to the degree distribution,
\begin{equation}
\label{pk2}
p(k) =C\frac{(Q_k+{\rm e}^{-\mu})^k}{k!} \exp \left[-\beta f(k)\right]  \ .
\end{equation}
where $C$ is a normalization constant. For example, a 
power-law tail in $p(k)$ may be generated by a suitable
choice of the weights $f(k)$ but it is not generic. 

A more detailed account of networks with generic local
interactions will be published elsewhere~\cite{pre}. Here
we turn to the simplest Hamiltonian with local interactions 
producing nontrivial optimized networks, see Fig.~1. It 
has the form
$\H_L = \H_1 + \H_2$ with
$\H_1 (\c) =-(1/2) \sum_i k_i^2 + \eta \sum_i k_i^3$
and
$\H_2 (\c) = \zeta \sum_{i<j}\delta_{k_i,1} c_{ij}
\delta_{k_j,1}$.
The first term
$(1/2) \sum_i k_i^2=(1/2) \sum_{ijk} c_{ik}c_{kj}$ gives the
number of paths of length two on the graph. It rewards
the formation of hubs, i.e. highly connected vertices, which in turn
lead to short distances. In fact this term has the maximally compact,
starlike configuration as its ground state. The collapse to a star, 
where the connectivity of the central vertex scales with the size 
of the graph, however, is  prevented 
by the regularization term $\eta \sum_i k_i^3$. 
The correlation term $\H_2$ with $\zeta \to \infty$ suppresses single, 
isolated links connecting two vertices of connectivity $1$. 
Without this term an extensive number of isolated links remains 
even in the limit $\beta \to \infty$ leading to graphs 
with a large number of disconnected parts.
A minimal connectivity of $1$ of each vertex is enforced. 
For this Hamiltonian, the free energy
(\ref{ZQ}) contains only one non-zero order parameter $Q_1$ 
given by 
$
p(1)=\kappa \frac{Q_1}{Q_1-e^{-\mu} } \ . 
$
The chemical potential $\mu$ is determined by $\sum_k k p(k)=\kappa$. 
Remarkably, the connectivity correlation can be obtained from $\H_2$ and 
single-vertex quantities, 
\begin{equation}
\label{q2}
q(k,k') \sim k p(k) t_k k' p(k') t_{k'} \exp \left[ -\beta g(k,k')\right] \ .
\end{equation}
The constant of proportionality is fixed by normalization and the 
$t_k$ are determined by the marginal distribution 
$\sum_{k'} q(k,k') = k p(k)/\kappa$, giving $t_1=\kappa/(T(\kappa-p(1)))$ and 
$t_k=T=\sqrt{1-p(1)^2/(\kappa-p(1))^2}$ for $k>1$. 

The properties of optimized networks resulting from the Hamiltonian 
$\H_L$ are readily inferred
from Eqs.~(\ref{pk2}) and~(\ref{q2}). 
At finite values $\beta$ one finds that the degree distribution 
(\ref{pk2}) has an exponentially decaying tail. 
In order to analyze the limit 
$\beta \to \infty$, we replace the sum over $k$ in (\ref{ZQ}) by an 
integral. One finds that the vertices arrange themselves 
into {\em hubs} of connectivity 
\begin{equation}
\label{hubconn}
k^* = (1 - 2 \eta)/4\eta
\end{equation}
and peripheral
vertices of connectivity $1$. The peripheral vertices are
connected only to hubs, while the hubs form an uncorrelated
random network.

A remarkably similar structure is found for compactified
networks generated by the Hamiltonian with nonlocal interactions 
$\H = H_1 + \lambda H_d$ with $H_d$
given by (\ref{Hd}) and $\H_1 = \eta \sum_i k_i^3$.
For $\lambda > 0$, the nonlocal part $\H_d$ favors networks
with short distances, while $\H_1$ prevents the collapse to
a star as before. Hence, by choosing $\lambda = 2/(N
\log N)$, one obtains a well-defined thermodynamic limit
with the average distance between vertices scaling as $\log
N$. 
We have studied this ensemble, as well as the case
of local interactions $\H_L$, by a Monte-Carlo link dynamics.
Starting, for example, from an Erd\"os-Reny graph, randomly chosen links
are moved to previously unlinked vertex pairs with probability 
$p=\mbox{min}(1,\exp[-\beta \Delta {\cal H} ])$ where 
$\Delta {\cal H}$ denotes the corresponding change in the Hamiltonian.
The minimum degree of $1$ is enforced throughout, self-links
are excluded~\cite{unlabel}. No dependence on the initial
conditions has been found. 
For the local Hamiltonian we use a network
with $N=200$, $\kappa=2.4$, $\eta=0.03$, for the nonlocal Hamiltonian
we use $N=100$, $\kappa=2.4$, $\eta=0.001$.
We averaged over $100$ samples. 
Fig.~2
juxtaposes analytical and numerical results for local 
interactions on the left with 
numerical results for nonlocal interactions on the right. 
The connectivity distribution $p(k)$ shows the 
formation of high-connectivity hubs in both cases. 

\begin{figure}[t]
{\small a)}\vspace{-.3cm}
 \includegraphics[width=\linewidth]{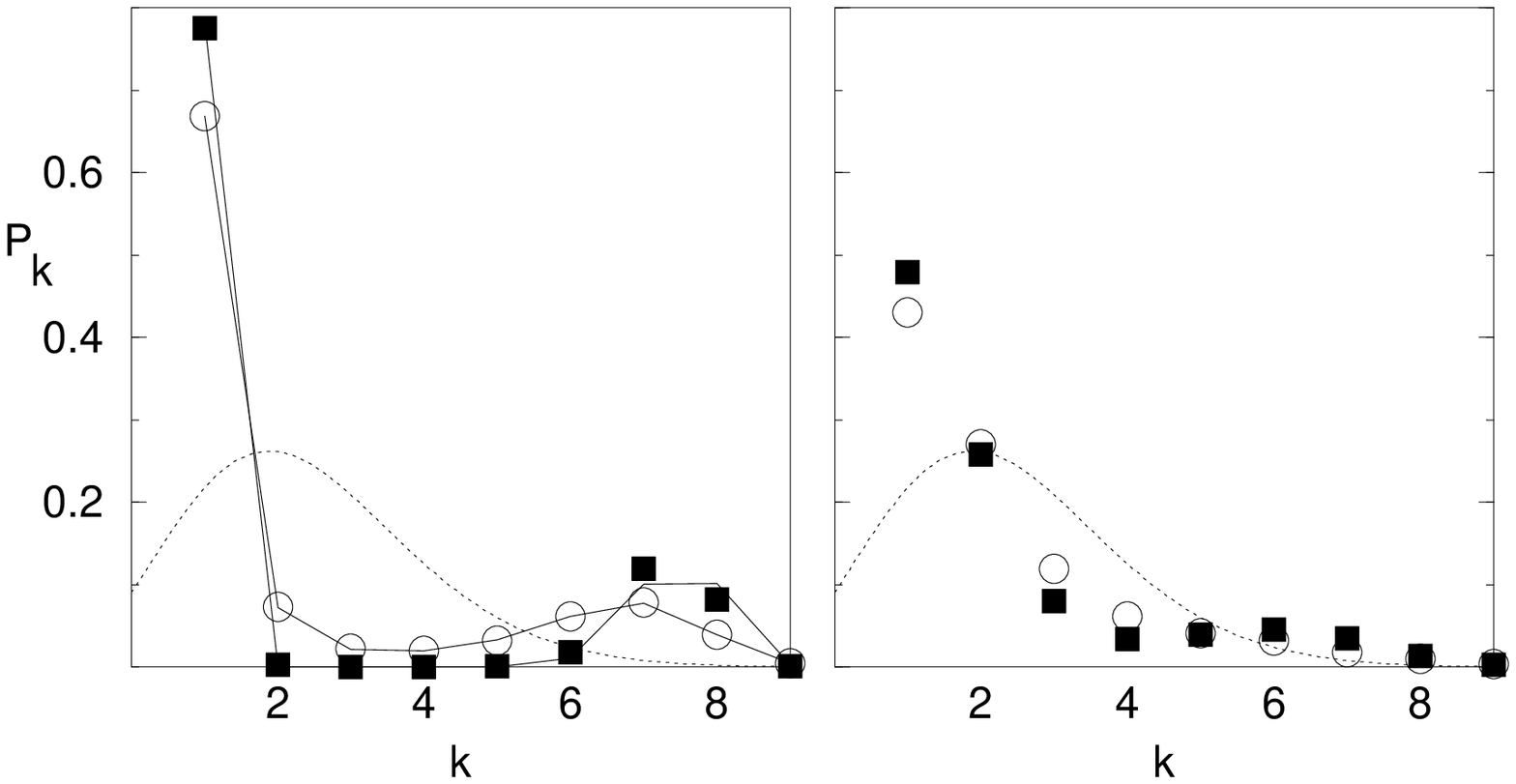}
{\small b)}\vspace{-.2cm}
\hspace*{.2 \textwidth}q(k,k')   \\
\hspace*{0.15cm}
\includegraphics[width=.48\linewidth]{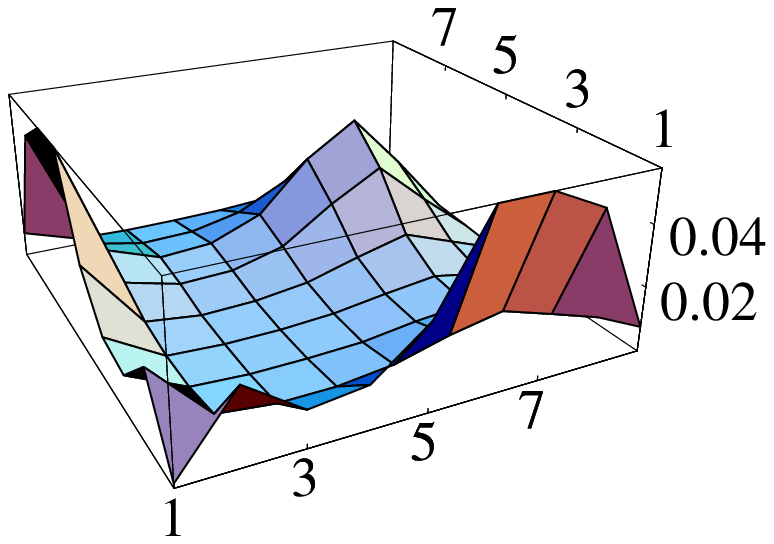}
\includegraphics[width=.48\linewidth]{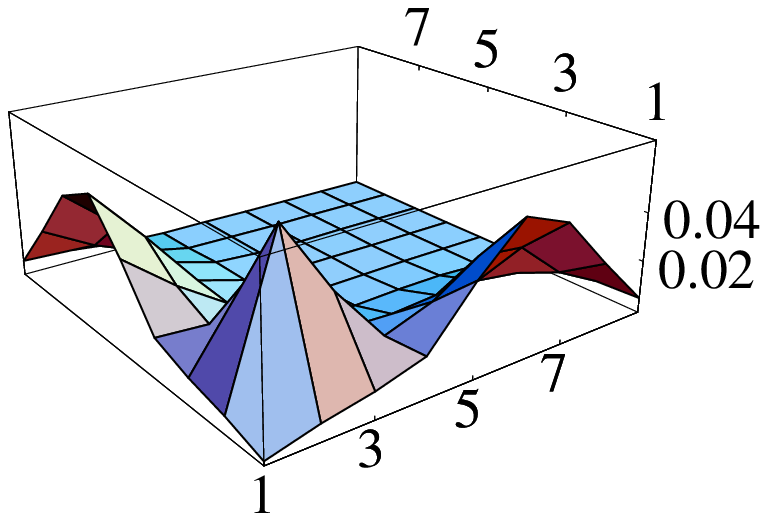}
{\small c)}\vspace{-.5cm}
\includegraphics[width=\linewidth]{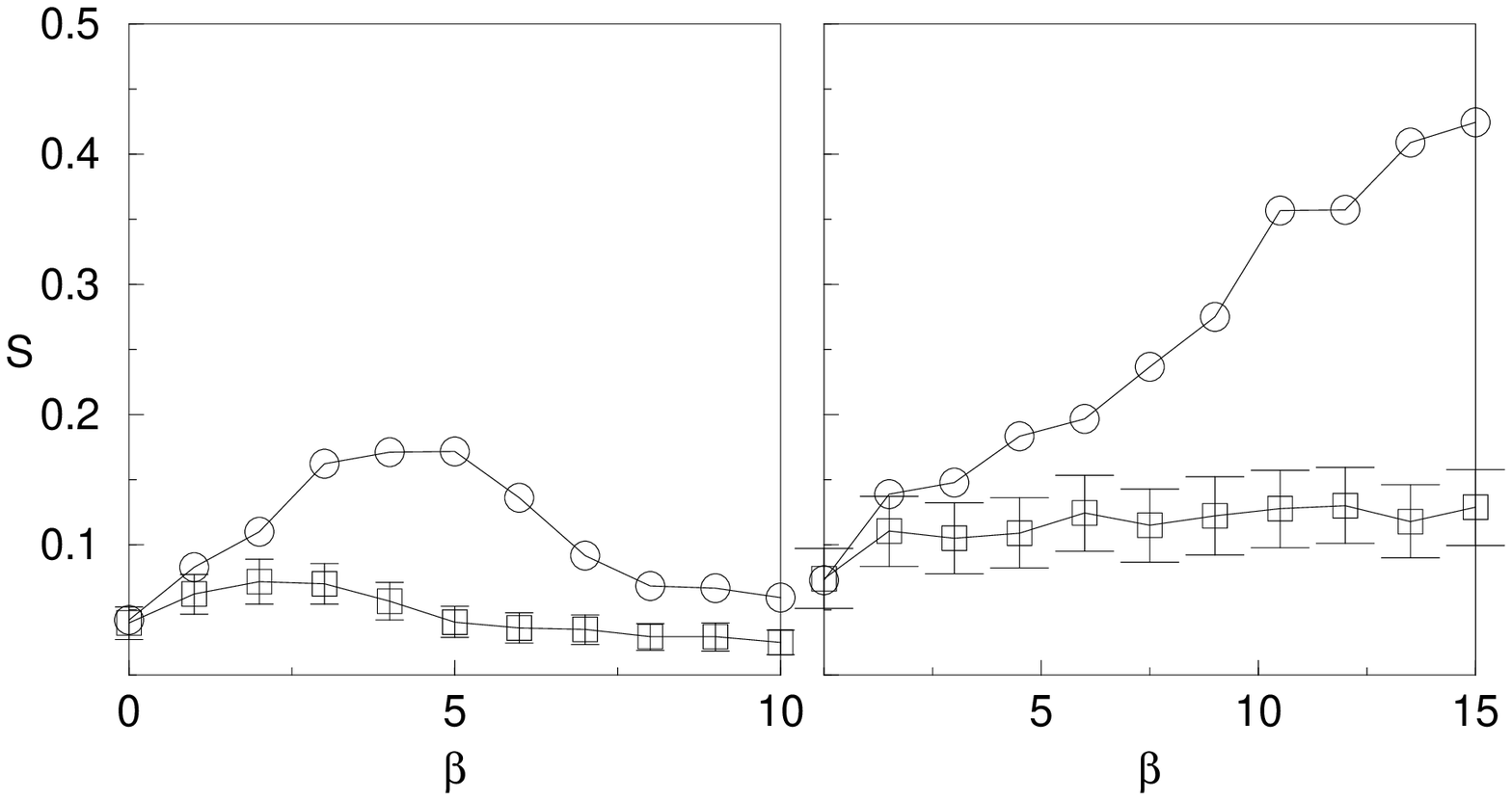}
{\small d)}\vspace*{-.5cm}
\includegraphics[width=\linewidth]{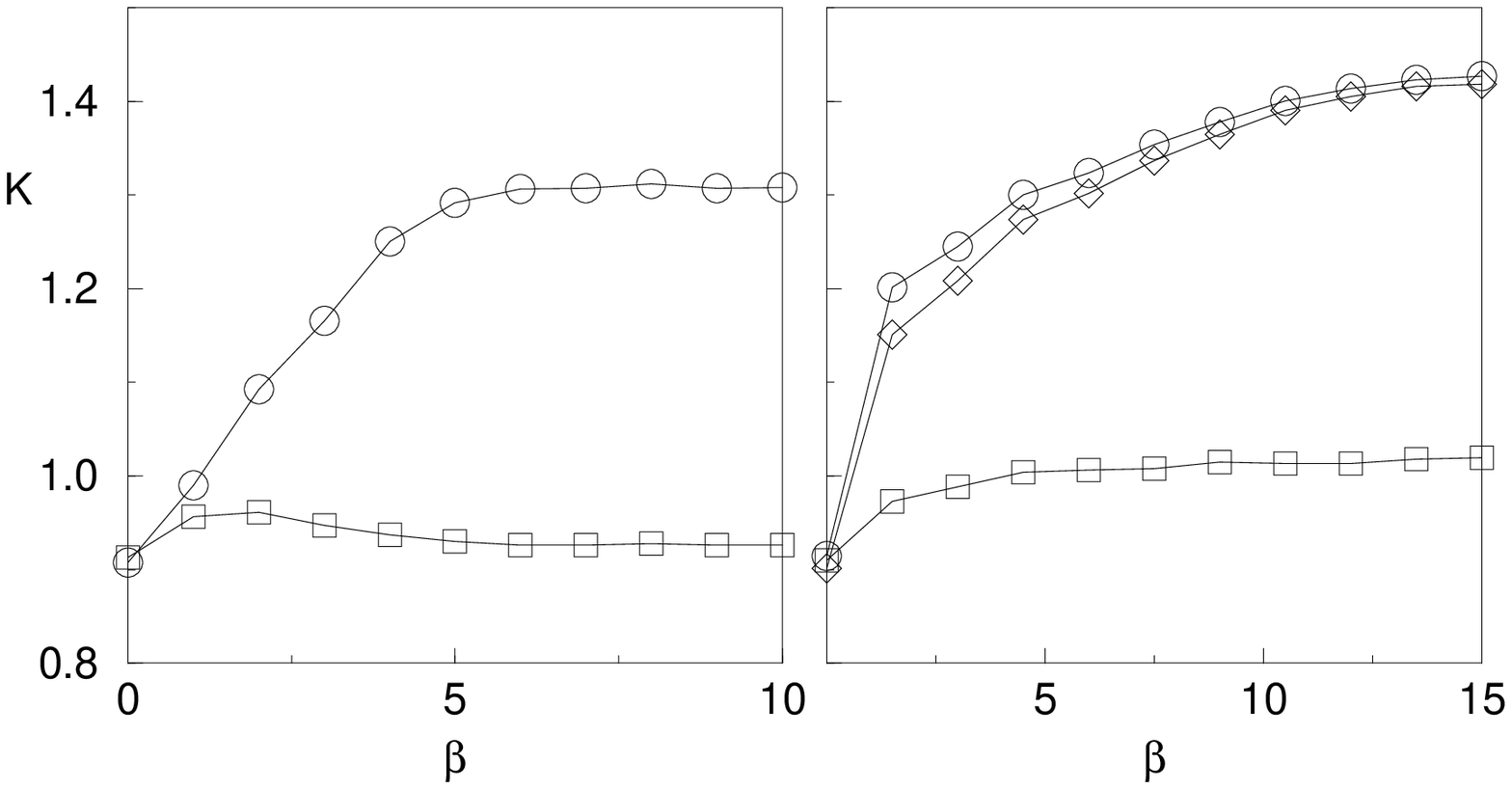}
\caption[]{\label{fig_all} Statistical features of correlated random
networks with local interactions (left) and nonlocal
interactions (right).

(a)~Single-point connectivity distribution $p(k)$ for various values
of~$\beta$. The Poisson form (dotted line),
data for intermediate $\beta$ (open circles)
and large $\beta$ (filled squares), and analytical values
(\ref{pk2}) for the case of local interactions (solid lines).

(b)~Neighbor connectivity distribution $q(k,k')$.
Left: Local interactions, analytical form (\ref{q2}) for
$\beta=3$. 
Right: Nonlocal interactions, numerical results for
$\beta=15$.

(c)~Average relative entropy
$\langle S(\hat q | \hat q_0)\rangle$ 
(circles), compared to the average sampling entropy 
$\langle S_0(\hat{q}_0)\rangle$ (squares) and its 
standard deviation $\langle \Delta S_0(\hat{q}_0)\rangle$, 
see text.

(d)~Average inverse distance $K$ as a
function of $\beta$ (open circles), compared to the same quantity for
the equivalent uncorrelated network (open squares) and
for the equivalent locally correlated network (right figure,
open diamonds); see text. 
}
\end{figure}

For local interactions the hub connectivity is given by 
(\ref{hubconn}), for non-local interaction the distribution 
remains broad even in the limit $\beta \to \infty$ (Fig.~2(a)) 
\cite{etas}. 
Low-connectivity vertices are preferentially attached to
hubs as indicated by the peaks at $q(1,k^*)$
for local interactions and the corresponding peaks of
$q(k,k')$ for nonlocal interactions (Fig.~2(b)).
The deviation from uncorrelated random networks is measured
by the {\em relative entropy}
$
S(q|q_0) = \sum_{k,k'} q(k,k') 
                       \log \,\frac{q(k,k')}{q_0(k,k')} \;.
$
In a {\it single} network of size $N$, we obtain the estimate 
$S(\hat q | \hat q_0)$ from the observed frequencies 
$\hat{q}(k,k')$ and $\hat{p}(k)$, with 
$\hat{q}_0 (k,k') = (k k'/\kappa^2) \hat{p}(k) \hat{p}(k')$. 
We then generate a sufficient number of uncorrelated random 
networks \cite{alg} with frequencies $\hat{q}^{\alpha}(k,k')$ and 
sampling entropies $S(\hat{q}^{\alpha}|\hat{q}_0)$; their average 
and standard deviation are denoted by $S_0(\hat{q}_0)$ and 
$\Delta S_0(\hat{q}_0)$ respectively. Connectivity correlations 
in the original network are significant if 
$S(\hat{q}|\hat{q}_0) - S_0(\hat{q}_0) \gtrsim \Delta S_0(\hat{q}_0)$. 
This is typically the case above a certain optimization degree $\beta$ on, 
as shown by the ensemble averages over $10$ samples 
$\langle S(\hat{q}|\hat{q}_0) \rangle$, 
$\langle S_0(\hat{q}_0)\rangle$, and $\langle \Delta S_0(\hat{q}_0)\rangle$ 
shown in figure 2c). 

Both kinds of networks become more compactified with
increasing $\beta$, as shown by the average inverse
distance $K \equiv (2/N(N-1)) \sum_{i<j} d_{ij}^{-1}$
(Fig.~2(d))~\cite{invdist}. We also plot $K$ for the
equivalent uncorrelated random networks; 
no such compactification is seen. Hence,
the one-point distribution $p(k)$ may miss important
functional properties. 
On the other hand,
the nonlocally interacting networks and their equivalent
locally interacting counterparts (constructed to have the
same $p(k)$ and $q(k,k')$) have a very similar degree of
compactification~\cite{alg}. 
This illustrates how {\em optimization induces
correlations}.

In summary, we have shown how interactions shape the
structure of a network. Hamiltonians beyond the
`single-vertex' form (\ref{H1}) generate
correlations such as a neighbor connectivity distribution
$q(k,k')$ which differs from that in uncorrelated networks.
Higher correlations can be defined in a similar
way~\cite{higher}. These correlations provide a more
detailed fingerprint of the interactions present than
the single-point connectivity distribution $p(k)$.
This observation should carry over to the dynamical rules
for non-equilibrium ensembles such as the
well-known growth models \cite{barabasi_review,newman_grow}.

In transcription control networks, structural motifs have
been identified that can be expressed in terms of
connectivity correlations~\cite{alon}. Such correlations have also
been observed in protein networks~\cite{maslov}.
In view of our findings for optimized networks, they appear
to be a natural consequence of the underlying dynamics and
functional optimization.
We expect the data to give important information on the
underlying design principles
of networks and on the selective forces governing their
evolution. Reverse engineering seems feasible, with the aim
of inferring the relevant dynamics from the data.
The {\em nonequilibrium} theory of correlated random networks
will thus be an important avenue for future research.

{\bf Acknowledgements}
Many thanks to Sergei Maslov for fruitful discussions and
for making \cite{maslov} available prior to publication.


\begin{references}

\bibitem{barabasi_review}
R. Albert and A.-L. Barab\'asi,
{\it Rev. Mod. Phys.} {\bf 47},74 (2002).

\bibitem{dorogovtsev_review}
S.N. Dorogovtsev and J.F.F. Mendes,
{\it Adv. Phys.}{\bf 51}, 1079 (2002).

\bibitem{gene_network}
See, e.g., www.mgs.bionet.nsc.ru/systems/mgl/genenet

\bibitem{protein_network}
P. Uetz {\it et. al.}, {\it Nature} {\bf 403},
623 (2000).

\bibitem{metabolic_network}
See, e.g.,
http://igweb.integratedgenomics.com/IGwit/.

\bibitem{bollobas}
B. Bollobas, {\it Random Graphs}
Academic Press, London, (1985).

\bibitem{unlabel}
The graphs can be defined with or without allowing
self-links (given by matrix elements $c_{ii} =1$). 
The corresponding ensembles differ only by corrections
of order $1/N$. Similarly, networks with unlabeled vertices
are rather simply related to the ensembles discussed here
in the large-$N$ limit; 
see~\cite{pre}.

\bibitem{newman}
M. E. J. Newman, S. H. Strogatz, and D. J. Watts,
{\it Phys. Rev.} {\bf E 64}, 026118 (2001).

\bibitem{burda}
Z. Burda, J.D. Correia, and A. Krzywicki,
{\it Phys. Rev.} {\bf E 64}, 046118 (2001).

\bibitem{higher}
The simplest higher term is ${\rm Tr} \, \c^3$, which counts
the number of triangles.

\bibitem{newman_grow}
D. S. Callaway {\it et. al.},
{\it Phys. Rev.} {\bf E 64}, 041902 (2001).

\bibitem{vespignani}
R. Pastor-Satorras, A. Vazquez, and A. Vespignani,
{\it Phys. Rev. Lett.}{\bf 87}, 258701 (2001).

\bibitem{alon}
S. S. Shen-Orr, et. al,{\it Nature Genetics} DOI:10.1038/ng881 (2002). 

\bibitem{maslov}
S. Maslov and K. Sneppen, {\it Science}, {\bf 296}, 
910-913 (2002).

\bibitem{logscal}
The same scaling is known for uncorrelated random networks;
see e.g. G. Szabo, M. Alava, and J. Kertesz, cond-mat/0203278 (2002).

\bibitem{smallworld}
A Hamiltonian of the form (\ref{Hd}) has been used to generate
small-world networks, see
N. Mathias and V. Gopal, {\it Phys. Rev.}{\bf E63}, 21117 (2001).

\bibitem{jeong}
H. Jeong, B. Tomber, R. Albert, Z.N. Oltvai,
and A.L. Barabasi, {\it Nature} {\bf 407}, 651 (2000).

\bibitem{sole}
R. Ferrer i Cancho and R. V. Sol\'e, cond-mat/0111222.

\bibitem{krzywicki}
A. Krzywicki, cond-mat/0110574.

\bibitem{tadpole}
This correspondence is strictly valid only in the
thermodynamic limit, where tadpole diagrams and multiple
contractions between the same pair of vertices can be
neglected.

\bibitem{pre}
J. Berg and M. L\"assig, to be published.

\bibitem{etas}
The parameters $\eta$ were chosen
such that the third moment of the degree distribution $\langle k^3 \rangle$ 
is the same for the two ensembles in the limit of large $\beta$.

\bibitem{invdist}
This form is chosen since it remains well-defined even for
disconnected components.

\bibitem{alg}
There is an efficient algorithm to produce an equivalent
uncorrelated random network from a given network: Choose
two edges connecting the pairs of vertices $i,j$ and $k,l$
at random with uniform probability and rewire them as $i,k$ and
$j,l$. Repeating this procedure sufficiently many times 
leaves $p(k)$ invariant but destroys all higher correlations.
It has been used in ref.~\cite{maslov} for protein
networks. Similarly, the equivalent locally correlated 
network is obtained by swapping only links 
with $k_i=k_l$ and $k_j=k_l$, which leaves
$p(k)$ and $q(k,k')$ invariant but destroys all higher
correlations.

\end{references}
\end{document}